\documentstyle[aps,preprint,tighten,floats,epsfig]{revtex}

\draft
\preprint{
\vbox{
\hbox{JLAB-THY-03-01}
}}

\begin{document}

\title{Higher Twists in the Pion Structure Function}
\author{W. Melnitchouk}
\address{Jefferson Lab,
	12000 Jefferson Avenue,
	Newport News, VA 23606}
\maketitle

\begin{abstract}
We calculate the QCD moments of the pion structure function using
Drell-Yan data on the quark distributions in the pion and a
phenomenological model for the resonance region.
The extracted higher twist corrections are found to be larger than those 
for the nucleon, contributing around 50\% of the lowest moment at
$Q^2=1$~GeV$^2$.
\end{abstract}

\vspace*{1cm}

%%%%%%%%%%%%%%%%%%%%%%%%%%%%%%%%%%%%%%%%%%%%%%%%%%%%%%%%%%%%%%%%%%%%%%%%

Understanding the structure of the pion represents a fundamental
challenge in QCD.
As the lightest $q\bar q$ bound state, the pion presents itself as 
somewhat of a dichotomy: on the one hand, its anomalously small mass 
suggests that it should be identified with the pseudo-Goldstone mode of 
dynamical breaking of chiral symmetry in QCD, essential for describing
the long-range structure and interactions of hadrons; on the other, high
energy scattering experiments reveal a rich substructure which can be
efficiently described in terms of current quarks and gluons.

The duality between quark and hadron degrees of freedom reveals itself
in most spectacular fashion through the phenomenon of Bloom-Gilman
duality in inclusive deep-inelastic scattering.
Here the inclusive $F_2$ structure function measured at low hadron final
state mass $W$, in the region dominated by low-lying resonances, has
been found in the case of the proton \cite{BG,NICU} to follow a global
scaling curve which describes high $W$ data, to which the resonance
structure function averages.
Furthermore, the equivalence of the averaged resonance and scaling
structure functions for each prominent resonance region separately
suggests that the resonance--scaling duality also exists to some
extent locally.

Within QCD, the appearance of Bloom-Gilman duality for the moments of
structure functions can be related through the operator product
expansion to the size of high twist corrections to the scaling
structure function \cite{RUJ}.
The apparent early onset of duality for the proton structure function
indicates the dominance of single-quark (leading twist) scattering even
at low momentum transfers.
It is not {\em a priori} clear, however, whether this is due to an
overall suppression of coherent effects in inclusive scattering, or
because of fortuitous cancellations of possibly large corrections.
Indeed, there are some indications from models of QCD that the workings
of duality may be rather different in the neutron than in the proton
\cite{IJMV,CI}, or for spin-independent and spin-dependent structure
functions.
Given that Bloom-Gilman duality is empirically established only for
baryons (specifically, the proton), while the application of theoretical
models is generally more straightforward in the meson sector, a natural
question to consider is whether, and to what extent, duality manifests
itself phenomenologically for the simplest $q\bar q$ system in QCD ---
the pion.

In this note we report the first analysis of the role of resonances
in the QCD moments of the pion structure function, and obtain a first
estimate of the size of higher twist corrections to the scaling
contribution.
Similar analyses for the nucleon have been made in
Refs.~\cite{JF2,LIUTI,JG1}.
According to the operator product expansion in QCD, at large $Q^2$ the
moments of the pion $F_2^\pi$ structure function,
\begin{eqnarray}
M_n(Q^2) &=& \int_0^1 dx\ x^{n-2}\ F_2^\pi(x,Q^2)\ ,
\label{MnDEF}
\end{eqnarray}
can be expanded as a power series in $1/Q^2$, with coefficients given by
matrix elements of local operators of a given twist $\tau$,
\begin{eqnarray}
M_n(Q^2) &=& \sum_{\tau=2}^\infty {\cal A}_{\tau-2}^n(\alpha_s(Q^2))
	\left( { 1 \over Q^2 } \right)^{\tau-2}\ .
\label{MnOPE}
\end{eqnarray}
Here the leading twist $\tau=2$ term ${\cal A}_0^n$ corresponds to free
quark scattering, and is responsible for the scaling of the structure
functions (modulo perturbative $\alpha_s(Q^2)$ corrections).
The higher twist terms ${\cal A}_{\tau>2}^n$ represent matrix elements
of operators involving both quark and gluon fields, and are suppressed
by powers of $1/Q^2$.
The higher twist contributions reflect the strength of nonperturbative
QCD effects, such as multi-parton correlations, associated with
confinement.

Note that the definition of $M_n(Q^2)$ includes the elastic contribution
at $W=m_\pi$, corresponding to $x = Q^2 / (W^2 - m_\pi^2 + Q^2) = 1$,
which is given by the square of the elastic pion form factor,
$F_\pi(Q^2)$,
\begin{eqnarray}
F_2^{\pi (\rm el)}(x=1,Q^2) &=&
2 m_\pi \nu\ \left( F_\pi(Q^2) \right)^2\ \delta(W^2 - m_\pi^2)\ .
\end{eqnarray}
Although negligible at high $Q^2$, the elastic contribution has been
found to be important numerically at intermediate $Q^2$ for moments of
the nucleon structure function \cite{JF2,LIUTI,JG1,EL,MEL}.
In Eq.~(\ref{MnDEF}) we use the Cornwall-Norton moments rather than the
Nachtmann moments, which are expressed in terms of the Nachtmann
scaling variable, $\xi = 2x/(1 + \sqrt{1+4x^2 m_\pi^2/Q^2})$, and
include effects of the target mass.
Because of the small value of $m_\pi$, the difference between the
variables $x$ and $\xi$, and therefore between the $x$- and
$\xi$-moments, is negligible for the pion.

The pioneering analysis of De~Rujula {\em et al.} \cite{RUJ} (see also
Ref.~\cite{JF2}) showed that the size of the higher twist matrix
elements directly governs the onset of quark-hadron duality.
Namely, there is a region of $n$ and $Q^2$ in which the moments of the
structure function are dominated by low mass resonances, where the
higher twist contributions are neither overwhelming nor negligible.
For example, even though there are large contributions from the
resonance region ($W \alt 2$~GeV) to the $n=2$ moment of the proton
$F_2$ structure function ($\sim 70\%$ at $Q^2=1$~GeV$^2$), the higher
twist effects are only of the order 10--20\% at the same $Q^2$ \cite{JF2}.
The question we address here is whether there is an analogous region for
the pion, where the resonance contributions are important, but higher
twist effects are small enough for duality to be approximately valid.

Of course, strictly speaking the distinction between the resonance
region and the deep inelastic continuum (DIS) is somewhat arbitrary, as
can be illustrated in the large $N_c$ limit of QCD.
There the final state in deep inelastic scattering from the pion is
populated by infinitely narrow resonances, even in the Bjorken limit,
while the structure function calculated at the parton level produces a
smooth, scaling function \cite{IJMV,EINHORN}.
Phenomenologically the spectrum of the excited states of the pion is
expected to be rather smooth sufficiently above the $\rho$ mass,
for $W \agt 1$~GeV.
Contributions from the excitation of heavier mesons are not expected to 
be easily discernible from the DIS continuum --- the $a_1$ meson, for
instance, appears at a mass $W \sim 1.3$~GeV, and has a rather broad
width ($\sim 350$--500~MeV).

The pion structure function has been measured in the $\pi N$ Drell-Yan
process \cite{NA3,NA10,E615} over a large range of $x$,
$0.2 \alt x \alt 1$, and for $Q^2$ typically $\agt 20$~GeV$^2$.
It has also been studied at HERA in semi-inclusive DIS at very low $x$
and high $W$ \cite{HERA}.
No data exist on $F_2^\pi$ at low $W$, however, in the region
where mesonic resonances would dominate the cross section.
The spectrum could in principle be reconstructed by observing low $t$
neutrons at low $W$ produced in the semi-inclusive charge-exchange
reaction, $e p \to e n X$, where $t$ is the momentum transfer squared
between the proton and neutron.
In the absence of such data, to obtain a quantitative estimate of the
importance of the resonance region, we model the pion spectrum at
low $W$ in terms of the elastic and $\rho$ pole contributions, on top
of the DIS continuum which is evaluated by evolving the leading twist
structure function to lower $Q^2$.
The latter can be reconstructed from parameterizations
\cite{GRVPI,SMRSPI,GRSPI} of leading twist quark distributions in the
pion obtained from global analyses of the pion Drell-Yan data.
In this work we use the low $Q^2$ fit from Ref.~\cite{GRVPI}, which
gives the leading twist parton distributions in the pion for
$Q^2 > 0.25$~GeV$^2$, although our conclusions do not change with the
use of other parameterizations \cite{SMRSPI,GRSPI}.
For the elastic contribution we use a parameterization of the world's
data \cite{MACK} which interpolates smoothly between the perturbative
QCD and photoproduction limits \cite{WMLONG}.

The $\rho$ contribution is described by the $\pi\to\rho$ transition
form factor, $F_{\pi\rho}(Q^2)$, and is expected to fall as $1/Q^4$
at large $Q^2$ (compared with $1/Q^2$ for $F_\pi(Q^2)$).
Since there is no empirical information on $F_{\pi\rho}(Q^2)$, we
consider several models in the literature, based on a relativistic
Bethe-Salpeter vertex function \cite{ITOGROSS}, a covariant
Dyson-Schwinger approach \cite{MARIS}, and light-cone QCD sum rules
\cite{KHODJ}.
These represent a range of $\sim 100\%$ in the magnitude of
$F_{\pi\rho}(Q^2)$ over the region of $Q^2$ covered in this analysis.
The calculation of Ref.~\cite{KHODJ} gives a somewhat smaller result
than in Refs.~\cite{ITOGROSS,MARIS}, which give a similar magnitude
for $F_{\pi\rho}$.
The difference between these can be viewed as an estimate of the
uncertainty in this contribution.

\begin{figure}[t]	% FIG 1
\begin{center}
\epsfig{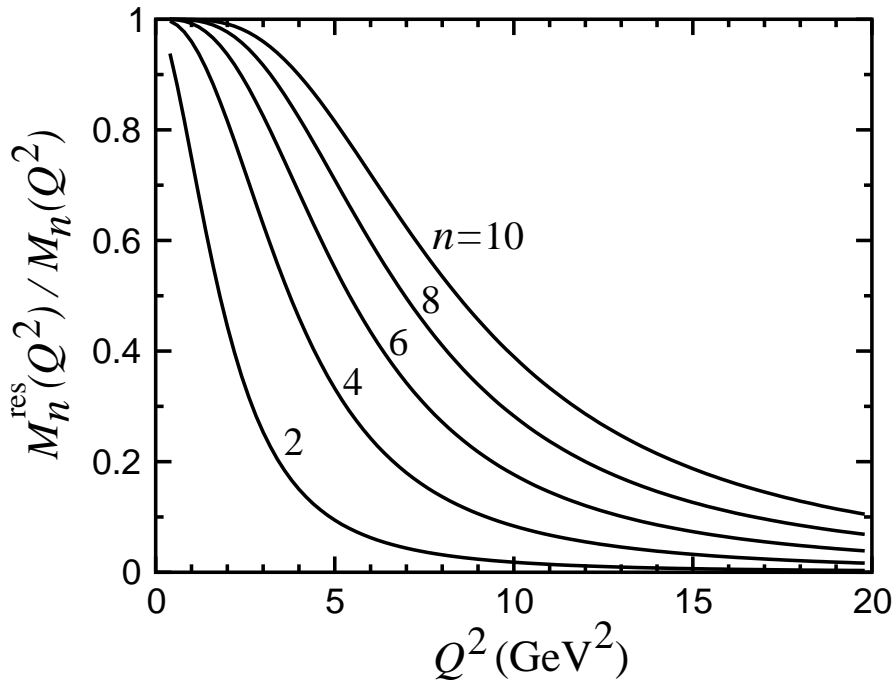}
\vspace*{0.5cm}
\caption{Contributions to moments of the pion structure function from
	the resonance region, $W < W_{\rm res} = 1$~GeV, relative to the 
	total.}
\end{center}
\end{figure}

In Fig.~1 we plot the contributions to the moments of the pion structure
function from the ``resonance region'', $M_n^{\rm res}(Q^2)$, as a ratio
to the total moment, for $n = 2, \cdots, 10$.
The resonance region here is defined by $W < W_{\rm res} \equiv 1$~GeV,
corresponding to restricting the integral in Eq.~(\ref{MnDEF}) to the
range $x_{\rm res} = Q^2/(W_{\rm res}^2 - m_\pi^2 + Q^2)$ to the
elastic point at $x=1$.
For the $n=2$ moment the low $W$ region contributes as much as 50\% at
$Q^2 = 2$~GeV$^2$, decreasing to $\alt 1\%$ for $Q^2 \agt 10$~GeV$^2$.
Higher moments, which are more sensitive to the large $x$ region,
subsequently receive larger contributions from low $W$.
The $n=10$ moment, for example, is virtually saturated by the resonance
region at $Q^2 = 2$~GeV$^2$, and still has some 40\% of its strength
coming from $W < 1$~GeV even at $Q^2 = 10$~GeV$^2$.

\begin{figure}[t]	% FIG 2
\begin{center}
\epsfig{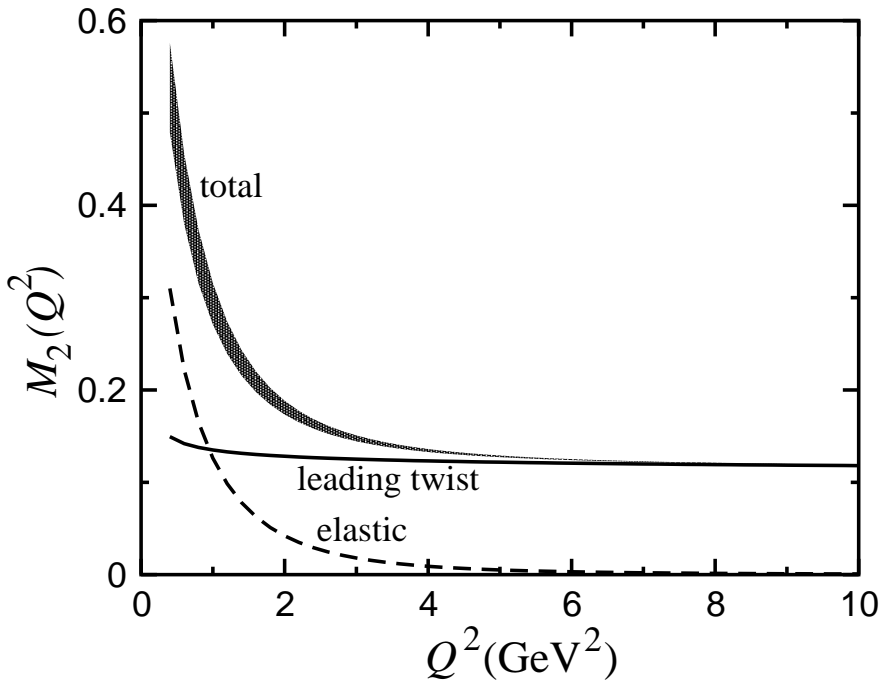}
\vspace*{0.5cm}
\caption{Lowest ($n=2$) moment of the pion structure function.
	The leading twist (solid) and elastic (dashed)
	contributions are shown, and the shaded region represents
	the total moment using different models for the $\pi\to\rho$
	transition.}
\end{center}
\end{figure}

The large size of the resonance contributions suggests that, at a given
scale $Q^2$, higher twist effects play a more important role in the
moments of the pion structure function than in the case of the nucleon.
The lowest ($n=2$) moment of $F_2^\pi$ is displayed in Fig.~2, including
the leading twist and elastic contributions to $M_2(Q^2)$.
The leading twist component,
\begin{eqnarray}
M_n^{\rm LT}(Q^2)
&=& \sum_q e_q^2\ \int_0^1 dx\ x^{n-1} q^\pi(x,Q^2)\ ,
\label{MnLT}
\end{eqnarray}
is expressed (at leading order in $\alpha_s(Q^2)$) in terms of the
twist-2 quark distributions in the pion, $q^\pi(x,Q^2)$.
The leading twist contribution is dominant at $Q^2 > 5$~GeV$^2$, while
the deviation of the total moment from the leading twist at lower $Q^2$
indicates the increasingly important role played by higher twists there.
In particular, while negligible beyond $Q^2 \approx 4$~GeV$^2$, the
elastic contribution is as large as the leading twist already at
$Q^2 \approx 1$~GeV$^2$.
The contribution from the $\pi\to\rho$ transition is more uncertain,
and the band in Fig.~2 represents the total moment calculated using
different models \cite{ITOGROSS,MARIS,KHODJ} of $F_{\pi\rho}(Q^2)$.
However, while the current uncertainty in this contribution is
conservatively taken to be $\sim 100\%$, doubling this would lead to a
modest increase of the band in Fig.~2.
Uncertainty arising from poor knowledge of the leading twist
distributions at small $x$ \cite{GRVPI,SMRSPI,GRSPI} is not expected
to be large.

\begin{figure}[t]	% FIG 3
\begin{center}
\epsfig{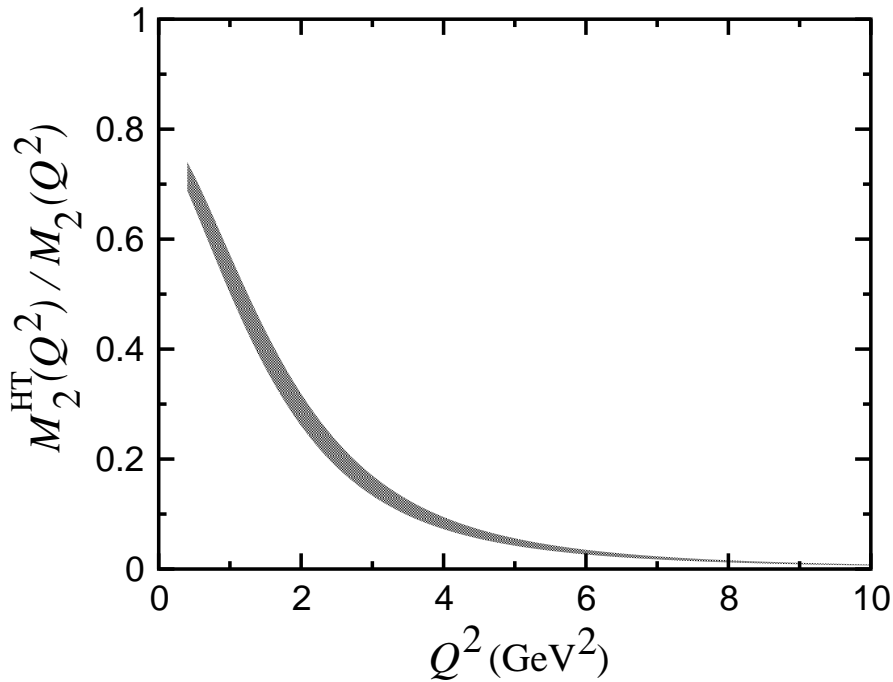}
\vspace*{0.5cm}
\caption{Higher twist contribution to the $n=2$ moment of the pion
	structure function, as a ratio to the total moment.
	The band indicates the uncertainty due to the model dependence
	of the $\pi\to\rho$ transition form factor.}
\end{center}
\end{figure}

The higher twist part of the moments can be defined as the difference
between the total moment $M_n(Q^2)$ and the leading twist contribution
in Eq.~(\ref{MnLT}), which includes a term arising from target mass
corrections,
\begin{eqnarray}
M_n^{\rm HT}(Q^2)
&=& M_n(Q^2) - M_n^{\rm LT}(Q^2) - M_n^{\rm TM}(Q^2)\ .
\end{eqnarray}
Although nonperturbative effects can in principle mix higher twist with
higher order effects in $\alpha_s$, rendering the formal separation of
the two ambiguous \cite{MUELLERHT} (the perturbative expansion itself
may not even be convergent), by restricting ourselves to the region of
$Q^2$ in which the $1/Q^2$ term is significantly larger than the next
order correction in $\alpha_s$, the ambiguity in defining the higher
twist terms can be neglected \cite{JF2}.
Because the target mass correction, $M_n^{\rm TM}(Q^2)$, which is formally
of leading twist, is proportional to $m_\pi^2/Q^2$, its contribution will
only be felt when $Q^2 \sim m_\pi^2$, which is far from the region where
the twist expansion is expected to be valid.
The higher twist contribution to the $n=2$ moment is plotted in Fig.~3
as a ratio to the total moment, as a function of $Q^2$.
The band represents an estimate of the uncertainty in the $\pi\to\rho$
transition form factor, as in Fig.~2.
The higher twist contribution is as large as the leading twist at
$Q^2 = 1$~GeV$^2$, is around 1/3 at $Q^2 = 2$~GeV$^2$, and vanishes
rapidly for $Q^2 \agt 5$~GeV$^2$.

The higher twist contribution at $Q^2 \sim 1$~GeV$^2$ appears larger
than that found in similar analyses of the proton $F_2$ \cite{JF2}
and $g_1$ \cite{JG1} structure functions.
This can be qualitatively understood in terms of the intrinsic transverse
momentum of quarks in the hadron, $\langle k_T^2 \rangle$, which typically
sets the scale of the higher twist effects \cite{RUJ,JF2,BB,GUNION}.
By analyzing the $x \to 1$ dependence of the measured $\mu^+\mu^-$ pairs
produced in $\pi N$ collisions, and the angular distribution at large $x$,
the E615 Collaboration \cite{E615} indeed finds the value
$\langle k_T^2 \rangle = 0.8 \pm 0.3$~GeV$^2$
within the higher twist model of Ref.~\cite{BB},
which is larger than the typical quark transverse momentum in the
nucleon (${\cal O}$(500~MeV)).
The implication is that duality would therefore be expected to set in
at larger $Q^2$ for the pion than for the nucleon.

In summary, we have evaluated moments of the pion structure function,
and studied in particular the role of the resonance region.
Making the reasonable assumption that the low $W$ resonant spectrum is
dominated by the elastic and $\pi\to\rho$ transitions, we have presented
a first quantitative estimate of the size of higher twist contributions.
For the lowest moment of $F_2^\pi$ we find that the resonance region
($W \alt 1$~GeV) contributes $\sim 50\%$ of the strength at
$Q^2 \approx 2$~GeV$^2$, dropping to below 10\% for $Q^2 \agt 5$~GeV$^2$.
The elastic component, while insignificant for $Q^2 \agt 3$~GeV$^2$, is
as large as the leading twist contribution at $Q^2 \approx 1$~GeV$^2$.
The higher twist corrections to the $n=2$ moment amount to $\sim 50\%$
at $Q^2 = 1$~GeV$^2$, $\sim 30\%$ at $Q^2 = 2$~GeV$^2$, and become
negligible beyond $Q^2 \approx 6$~GeV$^2$.

Uncertainties in these estimates are mainly due to the poor knowledge of 
the inclusive pion spectrum at low $W$, which limits the extent to which 
duality in the pion can be tested quantitatively.
Only the elastic form factor has been accurately measured to
$Q^2 \approx 2$~GeV$^2$, although at larger $Q^2$ it is poorly
constrained.
The inclusive pion spectrum can be extracted from data from the
semi-inclusive charge-exchange reaction, $e p \to e n X$, at low $t$,
for instance with CLAS at Jefferson Lab.
This could also allow one to determine the individual exclusive
channels at low $W$.
In addition, a Rosenbluth separation would allow the transverse and
longitudinal structure functions of the pion to be extracted.

%%%%%%%%%%%%%%%%%%%%%%%%%%%%%%%%%%%%%%%%%%%%%%%%%%%%%%%%%%%%%%%%%%%%%%%%
\acknowledgements

This work was supported by the U.S. DOE contract \mbox{DE-AC05-84ER40150},
under which the Southeastern Universities Research Association (SURA)
operates the Thomas Jefferson National Accelerator Facility (Jefferson
Lab).

%%%%%%%%%%%%%%%%%%%%%%%%%%%%%%%%%%%%%%%%%%%%%%%%%%%%%%%%%%%%%%%%%%%%%%%%


\references

\bibitem{BG}
E.~D.~Bloom and F.~J.~Gilman,
Phys. Rev. Lett. {\bf 25}, 1140 (1970);
%%CITATION = PRLTA,25,1140;%%
%
Phys. Rev. D {\bf 4}, 2901 (1971).
%%CITATION = PHRVA,D4,2901;%%
 
\bibitem{NICU}
I.~Niculescu {\it et al.},
Phys. Rev. Lett. {\bf 85}, 1182 (2000);
%%CITATION = PRLTA,85,1182;%%
%
{\it ibid} {\bf 85}, 1186 (2000).
%%CITATION = PRLTA,85,1186;%%

\bibitem{RUJ}
A.~De Rujula, H.~Georgi and H.~D.~Politzer,
Annals Phys. {\bf 103}, 315 (1977).
%%CITATION = APNYA,103,315;%%

\bibitem{IJMV}
N.~Isgur, S.~Jeschonnek, W.~Melnitchouk and J.~W.~Van Orden,
Phys. Rev. D {\bf 64}, 054005 (2001).
%%CITATION = HEP-PH 0104022;%%

\bibitem{CI}
F.~E.~Close and N.~Isgur,
Phys. Lett. B {\bf 509}, 81 (2001).
%%CITATION = HEP-PH 0102067;%%

\bibitem{JF2}
X.~Ji and P.~Unrau,
Phys. Rev. D {\bf 52}, 72 (1995).
%%CITATION = HEP-PH 9408317;%%

\bibitem{LIUTI}
C.~S.~Armstrong {\em et al.},
Phys. Rev. D {\bf 63}, 094008 (2001).
%%CITATION = HEP-PH 0104055;%%

\bibitem{JG1}
X.~Ji and P.~Unrau,
Phys. Lett. B {\bf 333}, 228 (1994);
%%CITATION = HEP-PH 9308263;%%
%
X.~Ji and W.~Melnitchouk,
Phys. Rev. D {\bf 56}, 1 (1997).
%CITATION = HEP-PH 9703363;%%

\bibitem{EL}
R.~Ent, C.~E.~Keppel and I.~Niculescu,
Phys. Rev. D {\bf 62}, 073008 (2000).
%%CITATION = PHRVA,D62,073008;%%

\bibitem{MEL}
W.~Melnitchouk,
Phys. Rev. Lett. {\bf 86}, 35 (2001);
%%CITATION = HEP-PH 0106073;%%
%
Nucl. Phys. A {\bf 680}, 52 (2000).
%%CITATION = NUPHA,A680,52;%%

\bibitem{EINHORN}
M.~B.~Einhorn,
Phys. Rev. D {\bf 14}, 3451 (1976).
%%CITATION = PHRVA,D14,3451;%%

\bibitem{NA3}
J.~Badier {\it et al.},
Z. Phys. C {\bf 18}, 281 (1983).
%%CITATION = ZEPYA,C18,281;%%

\bibitem{NA10}
B.~Betev {\it et al.},
Z. Phys. C {\bf 28}, 15 (1985).
%%CITATION = ZEPYA,C28,15;%%

\bibitem{E615}
J.~S.~Conway {\it et al.},
Phys. Rev. D {\bf 39}, 92 (1989).
%%CITATION = PHRVA,D39,92;%%

\bibitem{HERA}
G.~Levman,
J. Phys. G {\bf 28}, 1079 (2002).
%%CITATION = JPHGB,G28,1079;%%

\bibitem{GRVPI}
M.~Gluck, E.~Reya and A.~Vogt,
Z. Phys. C {\bf 53}, 651 (1992).
%%CITATION = ZEPYA,C53,651;%%

\bibitem{SMRSPI}
P.~J.~Sutton, A.~D.~Martin, R.~G.~Roberts and W.~J.~Stirling,
Phys. Rev. D {\bf 45}, 2349 (1992).
%%CITATION = PHRVA,D45,2349;%%

\bibitem{GRSPI}
M.~Gluck, E.~Reya and M.~Stratmann,
Eur. Phys. J. C {\bf 2}, 159 (1998).
%%CITATION = HEP-PH 9711369;%%

\bibitem{MACK}
H.~P.~Blok, G.~M.~Huber and D.~J.~Mack,
nucl-ex/0208011.
% Jefferson Lab Experiment E01-004

\bibitem{WMLONG}
W.~Melnitchouk,
hep-ph/0208258,
to appear in Eur. J. Phys. A.

\bibitem{ITOGROSS}
H.~Ito and F.~Gross,
Phys. Rev. Lett. {\bf 71}, 2555 (1993).
%%CITATION = PRLTA,71,2555;%%

\bibitem{MARIS}
P.~Maris and P.~C.~Tandy,
Phys. Rev. C {\bf 65}, 045211 (2002).
%%CITATION = NUCL-TH 0201017;%%

\bibitem{KHODJ}
A.~Khodjamirian,
Eur. Phys. J. C {\bf 6}, 477 (1999).
%%CITATION = HEP-PH 9712451;%%

\bibitem{MUELLERHT}
A.~H.~Mueller,
Phys. Lett. B {\bf 308}, 355 (1993).
%%CITATION = PHLTA,B308,355;%%

\bibitem{BB}
E.~L.~Berger and S.~J.~Brodsky,
Phys. Rev. Lett. {\bf 42}, 940 (1979).
%%CITATION = PRLTA,42,940;%%

\bibitem{GUNION}
J.~F.~Gunion, P.~Nason and R.~Blankenbecler,
Phys. Rev. D {\bf 29}, 2491 (1984).
%%CITATION = PHRVA,D29,2491;%%
\end{document}